\documentclass[journal=jacsat,manuscript=article]{achemso}
\usepackage{setspace}
\usepackage[utf8]{inputenc}
\usepackage[T1]{fontenc}
\usepackage{bm}
\usepackage{siunitx}
    \DeclareSIUnit\angstrom{\text {Å}}
    \DeclareSIUnit\bar{bar}
\author{Raphaël Salazar}
\affiliation{Synchrotron SOLEIL, L'Orme des Merisiers, Départementale 128, F-91190 Saint-Aubin, France}
\altaffiliation{Contributed equally to this work}
\email{raphael.salazar@synchrotron-soleil.fr}
\author{Sara Varotto}
\affiliation{Unité Mixte de Physique, CNRS, Thales, Université Paris-Saclay, 1 avenue Augustin Fresnel, 91767, Palaiseau France}
\altaffiliation{Contributed equally to this work}
\author{Céline Vergnaud}
\affiliation{Univ. Grenoble Alpes, CEA, CNRS, Grenoble INP, IRIG-SPINTEC, 38000 Grenoble, France}
\author{Vincent Garcia}
\affiliation{Unité Mixte de Physique, CNRS, Thales, Université Paris-Saclay, 1 avenue Augustin Fresnel, 91767, Palaiseau France}
\author{Stéphane Fusil}
\affiliation{Unité Mixte de Physique, CNRS, Thales, Université Paris-Saclay, 1 avenue Augustin Fresnel, 91767, Palaiseau France}
\author{Julien Chaste}
\affiliation{Université Paris-Saclay, CNRS, Centre de Nanosciences et de Nanotechnologies, 91120, Palaiseau, France}
\author{Thomas Maroutian}
\affiliation{Université Paris-Saclay, CNRS, Centre de Nanosciences et de Nanotechnologies, 91120, Palaiseau, France}
\author{Alain Marty}
\affiliation{Univ. Grenoble Alpes, CEA, CNRS, Grenoble INP, IRIG-SPINTEC, 38000 Grenoble, France}
\author{Frédéric Bonell}
\affiliation{Univ. Grenoble Alpes, CEA, CNRS, Grenoble INP, IRIG-SPINTEC, 38000 Grenoble, France}
\author{Debora Pierucci}
\affiliation{Université Paris-Saclay, CNRS, Centre de Nanosciences et de Nanotechnologies, 91120, Palaiseau, France}
\author{Abdelkarim Ouerghi}
\affiliation{Université Paris-Saclay, CNRS, Centre de Nanosciences et de Nanotechnologies, 91120, Palaiseau, France}
\author{François Bertran}
\affiliation{Synchrotron SOLEIL, L'Orme des Merisiers, Départementale 128, F-91190 Saint-Aubin, France}
\author{Patrick Le Fèvre}
\affiliation{Synchrotron SOLEIL, L'Orme des Merisiers, Départementale 128, F-91190 Saint-Aubin, France}
\author{Matthieu Jamet}
\affiliation{Univ. Grenoble Alpes, CEA, CNRS, Grenoble INP, IRIG-SPINTEC, 38000 Grenoble, France}
\email{matthieu.jamet@cea.fr}
\author{Manuel Bibes}
\affiliation{Unité Mixte de Physique, CNRS, Thales, Université Paris-Saclay, 1 avenue Augustin Fresnel, 91767, Palaiseau France}
\email{manuel.bibes@cnrs-thales.fr}
\author{Julien Rault}
\affiliation{Synchrotron SOLEIL, L'Orme des Merisiers, Départementale 128, F-91190 Saint-Aubin, France}

\title{Visualizing giant ferroelectric gating effects in large-scale WSe$_2$/BiFeO$_3$ heterostructures}


\begin{document}


\begin{abstract}
Multilayers based on quantum materials (complex oxides, topological insulators, transition-metal dichalcogenides, etc) have enabled the design of devices that could revolutionize microelectronics and optoelectronics. However, heterostructures incorporating quantum materials from different families remain scarce, while they would immensely broaden the range of possible applications. Here we demonstrate the large-scale integration of compounds from two highly-multifunctional families: perovskite oxides and transition-metal dichalcogenides (TMDs). We couple BiFeO$_3$, a room-temperature multiferroic oxide, and WSe$_2$, a semiconducting two-dimensional material with potential for photovoltaics and photonics. WSe$_2$ is grown by molecular beam epitaxy and transferred on a centimeter-scale onto BiFeO$_3$ films. Using angle-resolved photoemission spectroscopy, we visualize the electronic structure of 1 to 3 monolayers of WSe$_2$ and evidence a giant energy shift as large as 0.75 $\si{\eV}$ induced by the ferroelectric polarization direction in the underlying BiFeO$_3$. Such a strong shift opens new perspectives in the efficient manipulation of TMDs properties by proximity effects.
\end{abstract}

Transition metal dichalcogenides (TMDs) constitute a class of materials that have gathered a tremendous interest from the solid-state physics community focusing on two-dimensional (2D) materials. Schematically, TMDs of general formula MX$_2$ present a X-M-X configuration, where a metal is covalently bonded to two chalcogens (M = Mo, W; X = S, Se, Te). In TMDs, the global structure is made of several MX$_2$ planes bonded to one another by van der Waals (vdW) interactions. They hold promise for numerous applications in photonics and electronics \cite{manzeli_2d_2017,chen_gate-free_2018} owing to their large exciton binding energies \cite{ye_probing_2014,park_direct_2018,liu_direct_2019}, a theoretical ambipolarity originating from the X-M-X stacking and a transition from an indirect to a direct band gap when the material reaches the monolayer limit \cite{jin_direct_2013,zhao_evolution_2013}. They also exhibit a strong spin-orbit coupling \cite{zhu_giant_2011,xiao_coupled_2012,sugawara_spin-_2015} and a nonlinear and anisotropic Rasbha spin splitting has been predicted by \textit{ab initio} calculations \cite{cheng_nonlinear_2016} which are promising properties for spintronics. Owing to their ultrathin character, their electronic properties are easily tunable by proximity effects or any external stimuli. The engineering of the electronic and spin properties of TMDs is currently a very dynamical research area. Various methods based on strain engineering \cite{absor_polarity_2017}, hybrid TMDs heterostructures, hybrid dimension (2D/3D) heterostructures, or the exploitation of the twist degree of freedom have been reported to substantially modify their band structure \cite{yeh_direct_2016,kumari_engineering_2020,scuri_electrically_2020,zhang_flat_2020}.

Until recently, most studies on TMDs have focused on micrometer-sized flakes exfoliated from bulk samples \cite{novoselov_nobel_2011,kang_high-mobility_2015}. Extensive efforts are beginning to enable the growth of wafer-scale TMD single-crystal films \cite{wang_dual-coupling-guided_2022,dau_beyond_2018,pierucci_evidence_2022}. Yet, a remaining challenge is the transfer of these films onto wafers based on functional materials, notably from different structural and chemical families. A particularly appealing direction consists in coupling TMDs with electrically polarized materials such as ferroelectrics. This would enable a remanent modulation of their electronic properties and pave the way towards non-volatile memory devices such as ferroelectric field effect transistors (FeFET) with giant OFF/ON ratios. 

Since most technologically-relevant ferroelectrics are perovskite oxides such as BaTiO$_3$ (BTO), Pb(Zr, Ti)O$_3$ (PZT) and BiFeO$_3$ (BFO), there have been attempts to combine TMDs with such materials, albeit almost exclusively using exfoliated flakes. Several experiments showed a dependence of the photoluminescence (PL) response on the ferroelectric polarization direction in MoS$_2$/PZT \cite{lipatov_nanodomain_2019} or  MX$_2$/BTO \cite{li_polarization-mediated_2017,li_optical_2018,silva_resistive_2017,wen_ferroelectric-driven_2019}. Chen \textit{et al.} also defined a p-n homojunction on micron-sized WSe$_2$ flakes deposited on a BFO film with pre-poled up and down polarization domains \cite{chen_gate-free_2018}. However, these studies could not bring insight into the influence of ferroelectricity on the electronic structure of the TMD. 

In this work, we present a direct measurement of the modulation of the electronic band structure of a TMD by ferroelectric polarization using angle-resolved photoelectron emission spectroscopy (ARPES). We work with high-quality WSe$_2$ grown by molecular beam epitaxy (MBE) and transferred using chemical methods onto  BFO films grown by pulsed laser deposition (PLD) with different as-grown polarization states. First, we demonstrate that the single crystalline character of WSe$_2$ layers are preserved after transfer on BFO. Then, we show that 1 to 3 monolayers of WSe$_2$ exhibit a giant rigid band shift (\textit{ca.} 750 $\si{\meV}$ for a trilayer) when deposited on upward vs downward polarized BFO. This unprecedented value offers new opportunities to manipulate the electronic properties of TMDs by proximity effects.


\begin{figure}[!h]
    \centering
    \includegraphics{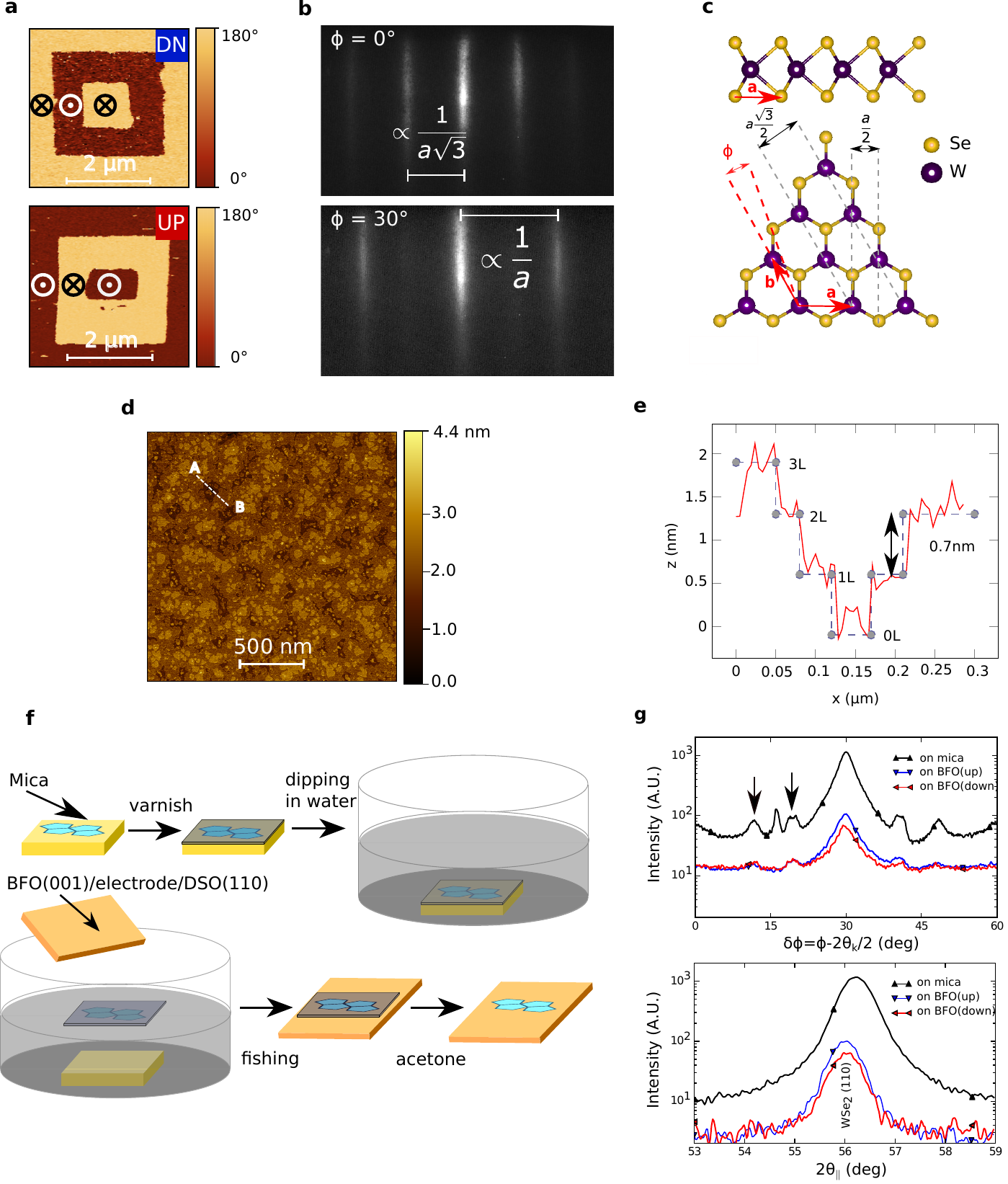}
    \caption{(a) PFM phase of DOWN BFO grown on SRO (blue square) and UP BFO grown on LSMO (red square). White arrows represent the UP out-of-plane electric polarization, black arrows represent the DOWN out-of-plane electric polarization.
(b) RHEED patterns of 3L WSe$_2$ along the directions [210] (azimuth $\phi$ = 30$^\circ$) and [100] (azimuth $\phi$ = 0$^\circ$).
(c) AFM image of MBE grown 3L WSe$_2$.
(d) AFM depth profile of 3L WSe$_2$ taken between the points \textbf{A} and \textbf{B} of the AFM image. The layers are 0.7 \si{\nm} thick. 
(e) Crystal structure of WSe$_2$, the vector $\mathbf{a}$ is the [100] direction, the vector $\mathbf{b}$ is the [010] direction. The violet balls represent the W atoms and the yellow balls the Se. The in-plane lattice parameter $a$ is indicated.
(f) Diagram of the transfer of MBE grown WSe$_2$ onto BFO (reproduced from \cite{dau_van_2019})
(g) \emph{Top}: azimuthal XRD intensity in log-scale before and after transfer on opposite polarized BFO substrates. The black arrows indicate the intensity from disoriented WSe$_2$ regions, invisible in linear scale.
\emph{Bottom}: In plane radial $\theta - 2\theta$ scan of XRD intensity in log-scale before and after transfer on opposite polarized BFO samples.}
    \label{Fig1}
\end{figure}


Epitaxial thin films of BFO (001) were prepared by PLD on (110)$_o$-oriented DyScO$_3$ (DSO) (the "o" subindex indicates orthorhombic notation) covered with bottom electrodes of SrRuO$_3$ (SRO) or La$_{0.7}$Sr$_{0.3}$MnO$_3$ (LSMO). Following Yu \textit{et al.}, \cite{yu_interface_2012} the choice of the bottom electrode sets the direction of the out-of-plane component of the ferroelectric polarization of the BFO film. Fig. \ref{Fig1}a presents piezoresponse force microscopy (PFM) phase images of the out-of-plane piezoresponse signal. In the virgin state, the contrast was homogeneous for both samples and could be switched reversibly by applying dc voltages higher than the coercive voltages (cf. concentric square patterns). Before switching, the piezoresponse occurred in phase with the excitation for the LSMO sample and with a phase shift of 180° for the SRO samples, indicating that the out-of-plane component of the polarization was pointing up and down, respectively.

To circumvent the temperature incompatibility for the direct growth of good quality WSe$_2$ films on BFO, we first grow WSe$_2$ films on mica substrates \cite{dau_van_2019,vergnaud_new_2020}. Reflection high-energy electron diffraction (RHEED) indicates the epitaxial growth on the single-crystalline mica substrate (Fig. \ref{Fig1}b). For the trilayer sample (3L), the atomic force microscopy (AFM) image of Fig. \ref{Fig1}c sample reveals a continuous coverage with 22\% of the surface corresponding to 3 ML (trilayer), 70\% to 2 ML (bilayer) and 8\% to 1 ML (monolayer). The area fraction of uncovered mica substrate (0L) is negligible. The line profile extracted on Fig. \ref{Fig1}d shows that the layers are 0.7 \si{\nm} thick as expected for bulk WSe$_2$. For the bilayer sample (2L), AFM images (see Supplementary Information) give a composition of 71\% 1 ML and 29\% 2 ML. 

The WSe$_2$ films were then wet-transferred on the BFO samples \cite{dau_van_2019}. In the case of the 3L sample, the whole surface was covered (cf. Fig. \ref{Fig1}f). X-ray diffraction before and after the transfer shows that the WSe$_2$ keeps its crystalline properties, i.e. the substrate does not induce additional strain or defects in the layer. The in-plane lattice parameter for WSe$_2$ on BFO is $a$ = 0.3284 \si{\nm}, compared to $a$ = 0.3272 \si{\nm} on mica. A value of $a$=0.3282 \si{\nm} for bulk WSe$_2$ is reported in Ref\cite{Schutte_1987}. This implies a 0.06\% strain for WSe$_2$ on BFO comparatively to a 0.3\% strain on mica, meaning that the  WSe$_2$ layer is nearly relaxed on BFO. Azimuthal X-ray diffraction demonstrates that the WSe$_2$ was grown with little angular dispersion and was transferred in exact conformity as only a little fraction of misoriented grains indicated by black arrows are detected before and after the operation as shown in Fig. \ref{Fig1}g (note that the diffraction intensity is in log-scale). PFM also confirmed that the BFO keeps its same polarization orientation after being covered by WSe$_2$ (see Supplementary Information). 


\begin{figure}[!h]
    \centering
    \includegraphics{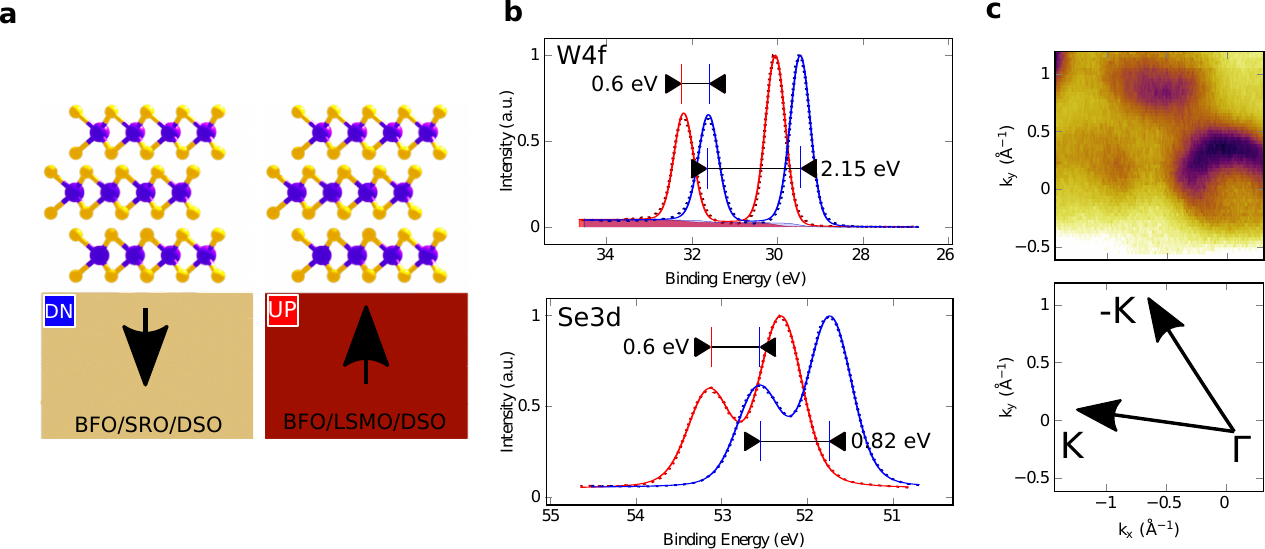}
    \caption{(a) Schematics of the 2D/3D system under study. Left: 3L WSe$_2$ transferred on down polarized BFO and right: 3L WSe$_2$ tranferred on up polarized BFO. In the following DOWN BFO is marked with a blue square and UP BFO by a red square. (b) XPS of 3L WSe$_2$ on DOWN BFO (blue) and UP BFO (red) showing the W 4f (top) and Se 3d (bottom) core levels. The photon energy was 500 eV. (c) Top: constant energy slice of ARPES data for 3L WSe$_2$ on DOWN BFO at E - E$_F$ = -0.89 eV. Bottom: second derivative of the slice.
}
    \label{Fig2}
\end{figure}

The first two stacks studied by ARPES are schematically shown in Fig. \ref{Fig2}a and consist in 3L WSe$_2$ transferred on DOWN and UP BFO respectively. X-ray photoemission spectroscopy (XPS) of We 4f and Se 3d core levels are displayed in Fig. \ref{Fig2}b. For both, we observe a single pair of peaks: 4f$_{7/2}$ and 4f$_{5/2}$ for W and 3d$_{5/2}$ and 3d$_{3/2}$ for Se with no evidence of secondary phase like WO$_x$ or SeO$_x$. Moreover the spin-orbit splittings (2.15 eV for W and 0.82 eV for Se) are in good agreement with tabulated value \cite{refXPS}. The only difference between DOWN and UP BFO polarities is a rigid energy shift of core level peaks by -0.6 eV which is characteristic of the electrostatic doping effect. In Fig. \ref{Fig2}c, the constant energy contours 500 \si{\meV} below the valence band maximum (VBM) show the two opposite valleys K and -K as well as the $\Gamma$ zone center at the bottom right of the image. WSe$_2$ clearly exhibits a single crystal orientation. At this energy, the lowest $\Gamma$ band already deviates from the radial symmetry to conform to an hexagonal one. The two nested bands at K and -K show the characteristic trigonal warping due the crystal symmetry. We conclude that the 3L WSe$_2$ transfer on BFO fully preserves the electronic structure and single crystalline character of WSe$_2$ grown by MBE.

\begin{figure}[!h]
    \centering
    \includegraphics{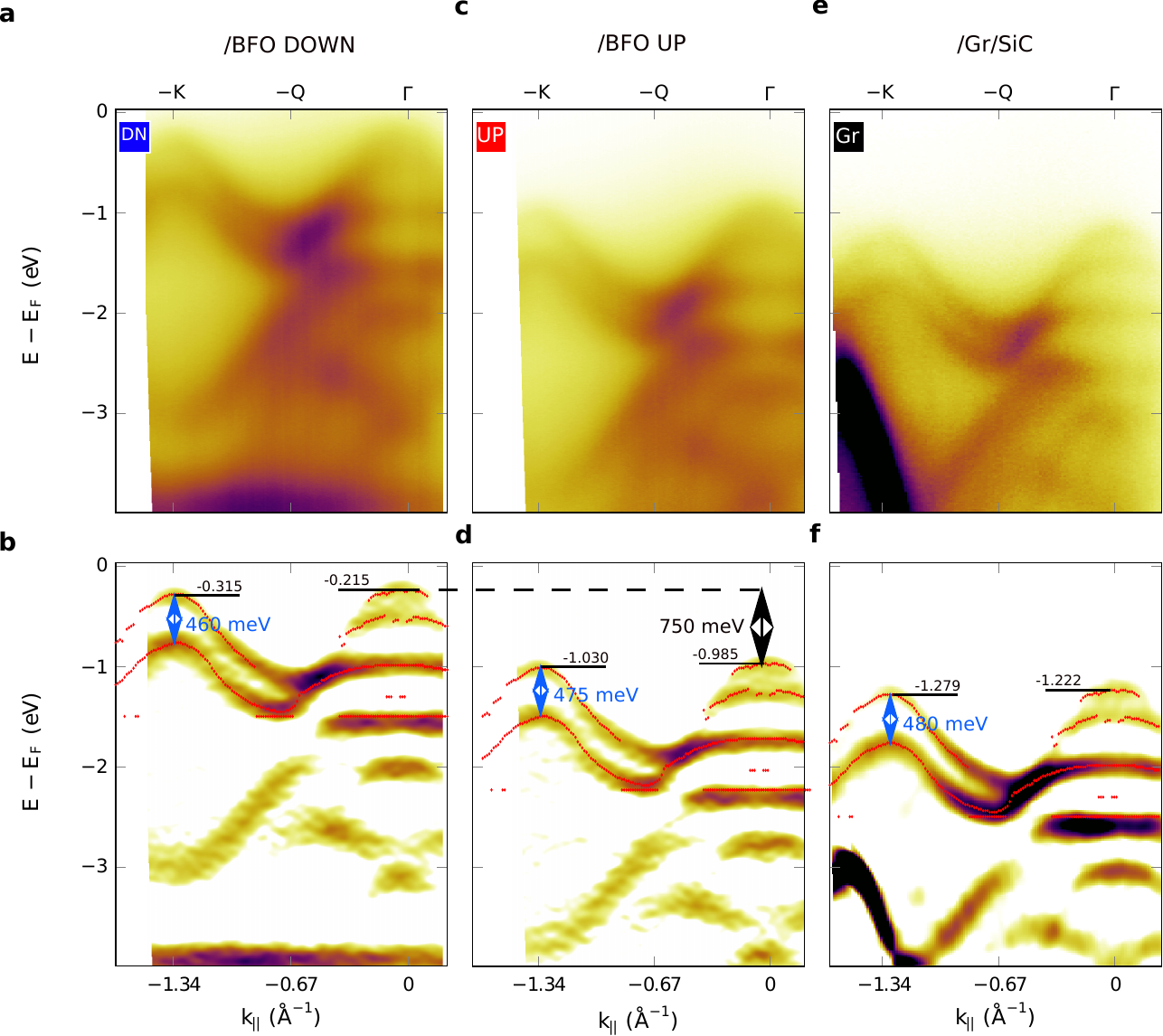}
    \caption{
    (a)  ARPES raw data and (b) second derivative of $\Gamma - K$ slice of the valence band of 3L WSe$_2$ on DOWN BFO. (c) ARPES raw data and (d) second derivative of $\Gamma - K$ slice of the valence band of 3L WSe$_2$ on UP BFO. (e) ARPES raw data and (f) second derivative of $\Gamma - K$ slice of the valence band of 3L WSe$_2$ transferred on graphene/SiC.
}
    \label{Fig3}
\end{figure}

Figure \ref{Fig3} displays ARPES measurements of the 3L films transferred on BFO and on a reference graphene substrate. Fig. \ref{Fig3}a and \ref{Fig3}c present the $\Gamma$-K slices of 3L WSe$_2$ on UP and DOWN BFO showing the features of the band structure from the top of the valence band down to 4 $\si{\eV}$ below the Fermi energy. Considering AFM images, black the observed band structure corresponds to the superposition of those of 3 ML, 2 ML and 1 ML of WSe$_2$. As a comparison, we provide in Fig. \ref{Fig3}e the same ARPES data for 3L WSe$_2$ transferred on graphene/SiC. In the valence band we see the two spin-splitted bands at K joining and separating into a threefold band at $\Gamma$. One $\Gamma$ branch culminates to define the VBM. The general shape of these $\Gamma$-K slices corresponds to the band structure predicted and experimentally measured for WSe$_2$ \cite{nguyen_visualizing_2019}. After Laplacian normalization, the bands are greatly sharpened showing clearly the aforementioned characteristics (Fig. \ref{Fig3}b, \ref{Fig3}d and \ref{Fig3}f). The splitting at K is consistent with what can be found in the literature \cite{nguyen_visualizing_2019,zhang_electronic_2016,nakamura_spin_2020,riley_direct_2014} and equal to $\Delta_{S-O}=$460 $\si{\meV}$ for WSe$_2$ on DOWN BFO, 475 $\si{\meV}$ on UP BFO and 480 $\si{\meV}$ on graphene. Comparing the band shifts for the three samples 3L WSe$_2$ on DOWN BFO, UP BFO and Gr/SiC we find that the VBM is located -0.215 $\si{\eV}$, -0.985 $\si{\eV}$ and -1.222 $\si{\eV}$ below the Fermi energy, respectively. The overall band structure of 3L WSe$_2$ is the same on BFO and graphene (the additional band appearing in the bottom left of Fig. \ref{Fig3}f corresponds to the graphene Dirac cone) and the principal effect of ferroelectric polarization is a rigid shift of the band structure, expected from electron accumulation (UP) or depletion (DOWN) from the BFO. Owing to the similar Pauling electronegativities of the metals Mo and W as well as the chalchogens S and Se, we follow the argument in Ref.\cite{dai_large_2019}. On UP BFO, (BiO)$^+$ termination would transfer electrons to the WSe$_2$ since $\chi(\text{Bi}) < \chi(\text{Se})$ while on DOWN BFO (FeO$_2$)$^-$ termination would deplete the layer since $\chi(\text{Se}) < \chi(\text{O})$. We measure an average rigid shift between the two samples of $\sim 750$ $\si{\meV}$. This is the major result of this study since this shift value is significantly higher than the 450 $\si{\meV}$ step measured by Kelvin force microscopy in Ref. \cite{chen_gate-free_2018}, which attests of the better coupling between the ferroelectric and the TMD in our samples. We note that it is also close to the value predicted from density functional theory for MoS$_2$/BiFeO$_3$ \cite{dai_large_2019}.


\begin{figure}[!h]
    \centering
    \includegraphics{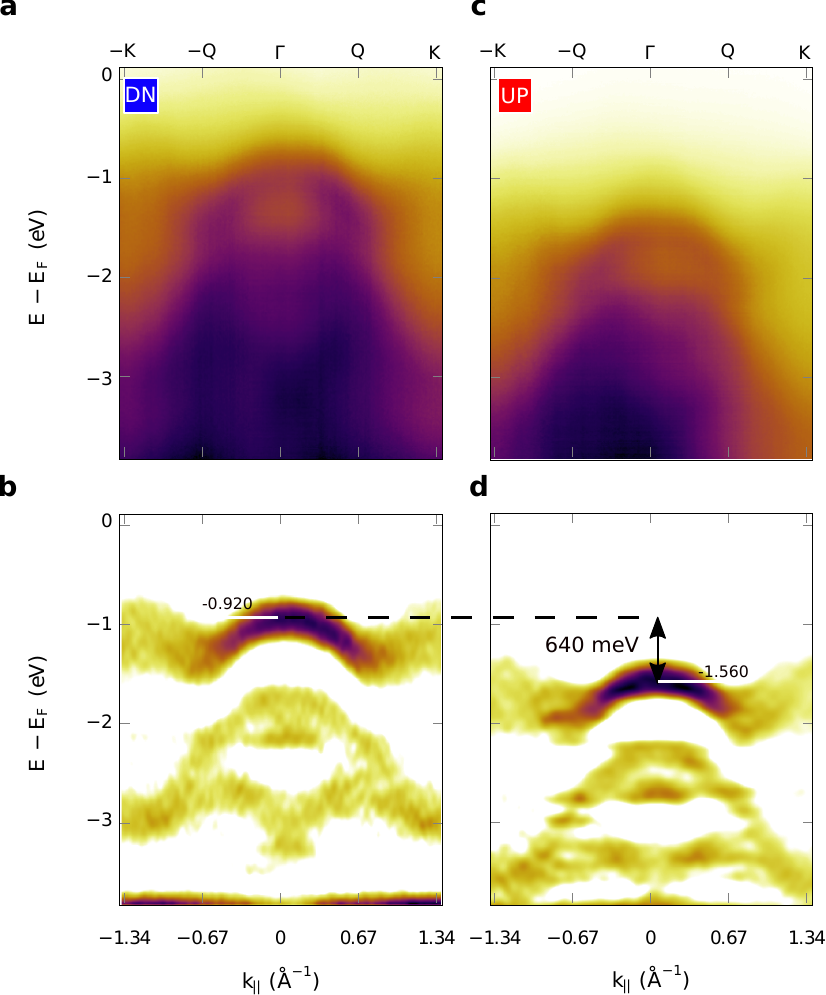}
    \caption{(a) ARPES raw data and (b) second derivative of $\Gamma - K$ slice of the valence band of 2L WSe$_2$ on DOWN BFO. (c) ARPES raw data and (d) second derivative of $\Gamma - K$ slice of the valence band of 2L WSe$_2$ on UP BFO. The ARPES was measured with h$\nu$ = 50 $\si{\eV}$ and linear horizontal polarization.
}
    \label{Fig4}
\end{figure}

Fig. \ref{Fig4} shows the ARPES measurements of 2L WSe$_2$ on DOWN and UP BFO in a $\Gamma$-K slice. As expected from AFM images, the data share similarities with the band structure of a standard 1L sample \cite{nguyen_visualizing_2019} in particular near $\Gamma$. A comparison between the data for the DOWN and UP samples again reveals a sizeable energy shift, here corresponding to 640 $\si{\meV}$, slightly lower than but consistent with the value found for the 3L sample, confirming the influence of the BFO polarization direction on the electronic structure. 
By enhancing the image contrast or changing the photon energy (see Supplementary Information), the ARPES data reveal features characteristic of 2 ML of WSe$_2$. In particular, a second $\Gamma$ band appears at higher energy. When comparing this additional band with the one reported in Ref. \cite{nguyen_visualizing_2019} for 2 ML of WSe$_2$, we notice that it appears flatter (corresponding to an apparent effective mass $m^\star=0.95 m_e$ vs. $m^\star=0.75 m_e$ in Ref. \cite{nguyen_visualizing_2019}). Interestingly, this flattening of the band (absent in the case of 3L sample) is also visible in Ref. \cite{nguyen_visualizing_2019} under the application of an electric field. One possible explanation for this band flattening could be the existence of a Rashba splitting induced in the WSe$_2$ by the BFO in our case and by the applied electric field in Ref. \cite{nguyen_visualizing_2019}. Indeed, Rashba spin-orbit coupling would split the bands along k which, in the absence of a high enough energy resolution (also due to the slight in-plane misorientation of the grains), would appear as a single, flatter band. Though beyond the scope of the present paper, this point would clearly deserve deeper investigation using for instance spin-resolved ARPES measurements.

In summary, the measurements on both 2L and 3L WSe$_2$/BFO large-scale heterostructures have proven that the band structure is rigidly shifted in dependence with the ferroelectric polarization. The atomic and electronic structures of WSe$_2$ remain essentially the same. We experimentally report the highest energy shift ever observed in TMD/3D ferroelectrics systems close to the theoretically predicted value. This giant shift is due to the formation of a dipole at the interface of the TMD and the ferroelectric since UP (DOWN) BFO presents positive (negative) charges at its surface. KPFM measurements have indeed shown that the work function of TMD flakes could be modulated by oppositely polarized domains
resulting in resistive switching behaviour \cite{chen_gate-free_2018,nguyen_toward_2015,li_polarization-mediated_2017}. Our results represent a clear step forward in the development of hybrid 2D/3D systems where proximity effects can efficiently modulate the electronic properties of the 2D material.

\vspace{2em}

\noindent \textbf{Growth of BFO thin-films by PLD.}
DOWN(UP) BFO samples were grown on DSO(110) substrates on top of a SRO(LSMO) electrode.
The SRO(LSMO) electrode was epitaxially grown to 5\si{\nm} thickness on the DSO substrate by PLD under a pressure of 0.2 \si{\milli\bar}(0.4 \si{\milli\bar}), a laser frequency of 2\si{\hertz} at 650\si{\degree}C.
The BFO was epitaxially grown to 40 \si{\nm} thickness on the electrode by PLD under a pressure of 0.4 \si{\milli\bar}, a laser frequency of 5 \si{\hertz} at 700\si{\degree}C.
The samples were then cooled slowly to room temperature (-10\si{\degree}C/min). For both polarization states the RMS roughness was typically under 1 \si{\nm} with visible atomic steps (not shown).

\noindent \textbf{Growth of WSe$_2$ films by MBE.}
The 15$\times$15 \si{\mm}$^2$ mica substrates from Ted Pella Inc. are first exfoliated mechanically using scotch tape to obtain a clean surface right before the introduction into the ultrahigh vacuum (UHV) growth chamber where the base pressure is kept in the low 10$^{-10}$ mbar. The substrates are then outgassed in the MBE chamber at 700$^{\circ}$C for 5 minutes to remove all the contaminants. The growth temperature as given by a thermocouple in contact with the sample holder is 920$^{\circ}$C \cite{vergnaud_new_2020}. WSe$_2$ layers are grown by co-evaporating W from an e-gun evaporator at a rate of 0.15 \AA /min and Se from an effusion cell. The Se partial pressure measured at the sample position is fixed at 10$^{-6}$ mbar. In-situ reflection high energy electron diffraction (RHEED) is used to monitor the WSe$_2$ crystal structure during the growth.

\noindent \textbf{Wet-transferring of the WSe$_2$ films.}
To transfer the WSe$_2$ layers from mica to BFO, we use a wet transfer method starting from the spin coating of a varnish on the sample. After evaporation of the solvent, the varnish film becomes solid and ensures the mechanical stability of the layers during the transfer process. The sample is then dipped into deionized water where the water penetrates at the interface between WSe$_2$ and mica. It progressively lifts off the varnish/WSe$_2$ stack which floats at the water surface. The floating stack is then gently fished using the target BFO substrate. The transferred layers are finally baked out on a hot plate at 80$^{\circ}$C for few minutes.

\noindent \textbf{X-ray diffraction.}
X-ray diffraction analysis was done with 5-axis Smartlab Rigaku diffractometer. The source was a Copper rotating anode ($K_{\alpha} =  1.54$ \si{\angstrom} operated at 45 \si{\kV} and 200 \si{\mA}). It is followed by a parabolic mirror and a parallel in-plane collimator resolving up to 0.5$\si{\degree}$. They constitute the primary optics, a second parallel collimator is used on the other side. All measurements were performed using a $K_{\beta}$ filter.

\noindent \textbf{Piezoresponse force microscopy.} The experiments were conducted with an atomic force microscope (Nanoscope V multimode, Bruker) and two external lockin detectors (SR830, Stanford Research) for the simultaneous acquisition of in-plane and out-of-plane responses. An external ac source (DS360, Stanford Research) was
used to excite the bottom electrode at a frequency of 35 kHz while the conducting Pt-coated tip was grounded. 

\noindent \textbf{Surface preparation for ARPES.}
The samples were mounted on Tantalum Omicron plates and electrically connected from the top with Molybdenum plates i.e. grounded to the manipulator. The samples stayed a few hours in UHV in order to degass as much surface contaminants as possible. We did not perform any kind of annealing prior to photoemission measurements for the WSe$_2$ on BFO samples to avoid damaging the BFO and its ferroelectric domain structure.

\noindent \textbf{Angle-resolved photoelectron emission spectroscopy.}
Unless specified otherwise we used horizontally polarized photons with energy h$\nu$ = 50 \si{\eV} at the Cassiopee beamline. The electron detector was a Scienta R4000 hemispherical Analyzer used at pass energy 20 \si{\eV}, with an angular resolution of 0.1$\si{\degree}$ and instrumental energy resolution of 18 \si{\meV}. The analyzer slits were vertical and curved. The Fermi energy $E_F$ was measured using a gold sample in equilibrium with the manipulator. The manipulator was at room temperature to prevent charging effects due to the semiconducting properties of both WSe$_2$ and BFO. We checked that the charging effect due to the beam flux was negligible. In the 3L sample, ARPES measurements correspond to the combination of the band structures of 3 ML (22\%), 2 ML (70\%) and 1 ML (8\%). Similarly, in the 2L sample, ARPES measurements correspond to the combination of the band structures of 2 ML (29\%) and 1 ML (71\%). The 3L and 2L labels were attributed on the basis of the number of visible bands at $\Gamma$, three for the 3L sample (cf. Fig. \ref{Fig3}) and two for the 2L sample (cf. Supplementary Information).

\noindent \textbf{ARPES data analysis.}
The ARPES data was all analysed using the homemade CASSIOpy software available at: \url{https://gitlab.com/SLZ_Raph/cassiopy/}. 
The band profiles were extracted iteratively from the EDC of the second derivatives of the measured ARPES slices by fitting a gaussian curve around the prominent peaks intensity points.
The second derivatives images were calculated taking into account inhomogeneous units or steps \cite{zhang_precise_2011} after an appropriate gaussian smoothing. We cut all the values below zero: $\nabla^2f$ < 0 $:=$ 0.
Large angular range slices are produced from a set of three acquisitions is measured with tilt angles -10,0,10\si{\degree} and merged into one spectrum. The overlapping regions are the average of all spectra at a given coordinate. Merging discontinuities are fully corrected with a linear correction.

\section*{Acknowledgements}

The authors acknowledge the financial support from the ANR CORNFLAKE (ANR-18-CE24-0015-01), ANR project MAGICVALLEY (ANR-18-CE24-0007), the LANEF framework (ANR-10-LABX-0051), the ERC AdG "FRESCO" (833973), and the EU H2020 Graphene Flagship.

\vspace{2em}

\noindent The authors declare no competing interests.

\vspace{2em}

\noindent Supporting information (additional X-ray diffraction, atomic force microscopy and ARPES data) is available free of charge via the internet at \url{http://pubs.acs.org}.

\pagebreak


\begin{thebibliography}{10}
\urlstyle{rm}
\expandafter\ifx\csname url\endcsname\relax
  \def\url#1{\texttt{#1}}\fi
\expandafter\ifx\csname urlprefix\endcsname\relax\def\urlprefix{URL }\fi
\expandafter\ifx\csname doiprefix\endcsname\relax\def\doiprefix{DOI: }\fi
\providecommand{\bibinfo}[2]{#2}
\providecommand{\eprint}[2][]{\url{#2}}

\bibitem{giustino_2021_2021}
\bibinfo{author}{Giustino, F.} \emph{et~al.}
\newblock \bibinfo{journal}{\bibinfo{title}{The 2021 quantum materials
  roadmap}}.
\newblock {\emph{J. Phys. Mater.}} \textbf{\bibinfo{volume}{3}},
  \bibinfo{pages}{042006} (\bibinfo{year}{2021}).

\bibitem{sando_bifeo_2014}
\bibinfo{author}{Sando, D.}, \bibinfo{author}{Barthélémy, A.} \&
  \bibinfo{author}{Bibes, M.}
\newblock \bibinfo{journal}{\bibinfo{title}{{BiFeO$_3$} epitaxial thin films
  and devices: past, present and future}}.
\newblock {\emph{Journal of Physics: Condensed Matter}}
  \textbf{\bibinfo{volume}{26}}, \bibinfo{pages}{473201},
  \doiprefix\url{10.1088/0953-8984/26/47/473201} (\bibinfo{year}{2014}).

\bibitem{kolobov_two-dimensional_2016}
\bibinfo{author}{Kolobov, A.~V.} \& \bibinfo{author}{Tominaga, J.}
\newblock \emph{\bibinfo{title}{Two-{Dimensional} {Transition}-{Metal}
  {Dichalcogenides}}}, vol. \bibinfo{volume}{239} of
  \emph{\bibinfo{series}{Springer {Series} in {Materials} {Science}}}
  (\bibinfo{address}{Cham}, \bibinfo{year}{2016}).

\bibitem{manzeli_2d_2017}
\bibinfo{author}{Manzeli, S.}, \bibinfo{author}{Ovchinnikov, D.},
  \bibinfo{author}{Pasquier, D.}, \bibinfo{author}{Yazyev, O.~V.} \&
  \bibinfo{author}{Kis, A.}
\newblock \bibinfo{journal}{\bibinfo{title}{{2D} transition metal
  dichalcogenides}}.
\newblock {\emph{Nature Reviews Materials}}
  \textbf{\bibinfo{volume}{2}}, \bibinfo{pages}{1--15},
  \doiprefix\url{10.1038/natrevmats.2017.33} (\bibinfo{year}{2017}).

\bibitem{chen_gate-free_2018}
\bibinfo{author}{Chen, J.-W.} \emph{et~al.}
\newblock \bibinfo{journal}{\bibinfo{title}{A gate-free monolayer {WSe$_2$} pn
  diode}}.
\newblock {\emph{Nature Communications}}
  \textbf{\bibinfo{volume}{9}}, \bibinfo{pages}{3143},
  \doiprefix\url{10.1038/s41467-018-05326-x} (\bibinfo{year}{2018}).

\bibitem{ye_probing_2014}
\bibinfo{author}{Ye, Z.} \emph{et~al.}
\newblock \bibinfo{journal}{\bibinfo{title}{Probing excitonic dark states in
  single-layer tungsten disulphide}}.
\newblock {\emph{Nature}} \textbf{\bibinfo{volume}{513}},
  \bibinfo{pages}{214--218}, \doiprefix\url{10.1038/nature13734}
  (\bibinfo{year}{2014}).

\bibitem{park_direct_2018}
\bibinfo{author}{Park, S.} \emph{et~al.}
\newblock \bibinfo{journal}{\bibinfo{title}{Direct determination of monolayer
  {MoS$_2$ and} {WSe$_2$ exciton} binding energies on insulating and metallic
  substrates}}.
\newblock {\emph{2D Materials}} \textbf{\bibinfo{volume}{5}},
  \bibinfo{pages}{025003}, \doiprefix\url{10.1088/2053-1583/aaa4ca}
  (\bibinfo{year}{2018}).

\bibitem{liu_direct_2019}
\bibinfo{author}{Liu, F.}, \bibinfo{author}{Ziffer, M.~E.},
  \bibinfo{author}{Hansen, K.~R.}, \bibinfo{author}{Wang, J.} \&
  \bibinfo{author}{Zhu, X.}
\newblock \bibinfo{journal}{\bibinfo{title}{Direct {Determination} of
  {Band}-{Gap} {Renormalization} in the {Photoexcited} {Monolayer} {MoS$_2$}}}.
\newblock {\emph{Physical Review Letters}}
  \textbf{\bibinfo{volume}{122}}, \bibinfo{pages}{246803},
  \doiprefix\url{10.1103/PhysRevLett.122.246803} (\bibinfo{year}{2019}).

\bibitem{jin_direct_2013}
\bibinfo{author}{Jin, W.} \emph{et~al.}
\newblock \bibinfo{journal}{\bibinfo{title}{Direct {Measurement} of the
  {Thickness}-{Dependent} {Electronic} {Band} {Structure} of {MoS$_2$} {Using}
  {Angle}-{Resolved} {Photoemission} {Spectroscopy}}}.
\newblock {\emph{Physical Review Letters}}
  \textbf{\bibinfo{volume}{111}}, \bibinfo{pages}{106801},
  \doiprefix\url{10.1103/PhysRevLett.111.106801} (\bibinfo{year}{2013}).

\bibitem{zhao_evolution_2013}
\bibinfo{author}{Zhao, W.} \emph{et~al.}
\newblock \bibinfo{journal}{\bibinfo{title}{Evolution of {Electronic}
  {Structure} in {Atomically} {Thin} {Sheets} of {WS$_2$} and {WSe$_2$}}}.
\newblock {\emph{ACS Nano}} \textbf{\bibinfo{volume}{7}},
  \bibinfo{pages}{791--797}, \doiprefix\url{10.1021/nn305275h}
  (\bibinfo{year}{2013}).

\bibitem{zhu_giant_2011}
\bibinfo{author}{Zhu, Z.~Y.}, \bibinfo{author}{Cheng, Y.~C.} \&
  \bibinfo{author}{Schwingenschlögl, U.}
\newblock \bibinfo{journal}{\bibinfo{title}{Giant spin-orbit-induced spin
  splitting in two-dimensional transition-metal dichalcogenide
  semiconductors}}.
\newblock {\emph{Physical Review B}}
  \textbf{\bibinfo{volume}{84}}, \bibinfo{pages}{153402},
  \doiprefix\url{10.1103/PhysRevB.84.153402} (\bibinfo{year}{2011}).

\bibitem{xiao_coupled_2012}
\bibinfo{author}{Xiao, D.}, \bibinfo{author}{Liu, G.-B.},
  \bibinfo{author}{Feng, W.}, \bibinfo{author}{Xu, X.} \& \bibinfo{author}{Yao,
  W.}
\newblock \bibinfo{journal}{\bibinfo{title}{Coupled {Spin} and {Valley}
  {Physics} in {Monolayers} of {MoS$_2$} and {Other} {Group}-{VI}
  {Dichalcogenides}}}.
\newblock {\emph{Physical Review Letters}}
  \textbf{\bibinfo{volume}{108}}, \bibinfo{pages}{196802},
  \doiprefix\url{10.1103/PhysRevLett.108.196802} (\bibinfo{year}{2012}).

\bibitem{sugawara_spin-_2015}
\bibinfo{author}{Sugawara, K.}, \bibinfo{author}{Sato, T.},
  \bibinfo{author}{Tanaka, Y.}, \bibinfo{author}{Souma, S.} \&
  \bibinfo{author}{Takahashi, T.}
\newblock \bibinfo{journal}{\bibinfo{title}{Spin- and valley-coupled electronic
  states in monolayer {WSe$_2$} on bilayer graphene}}.
\newblock {\emph{Applied Physics Letters}}
  \textbf{\bibinfo{volume}{107}}, \bibinfo{pages}{071601},
  \doiprefix\url{10.1063/1.4928658} (\bibinfo{year}{2015}).

\bibitem{cheng_nonlinear_2016}
\bibinfo{author}{Cheng, C.}, \bibinfo{author}{Sun, J.-T.},
  \bibinfo{author}{Chen, X.-R.}, \bibinfo{author}{Fu, H.-X.} \&
  \bibinfo{author}{Meng, S.}
\newblock \bibinfo{journal}{\bibinfo{title}{Nonlinear {Rashba} spin splitting
  in transition metal dichalcogenide monolayers}}.
\newblock {\emph{Nanoscale}} \textbf{\bibinfo{volume}{8}},
  \bibinfo{pages}{17854--17860}, \doiprefix\url{10.1039/C6NR04235J}
  (\bibinfo{year}{2016}).

\bibitem{absor_polarity_2017}
\bibinfo{author}{Absor, M. A.~U.} \emph{et~al.}
\newblock \bibinfo{journal}{\bibinfo{title}{Polarity tuning of
  spin-orbit-induced spin splitting in two-dimensional transition metal
  dichalcogenides}}.
\newblock {\emph{Journal of Applied Physics}}
  \textbf{\bibinfo{volume}{122}}, \bibinfo{pages}{153905},
  \doiprefix\url{10.1063/1.5008475} (\bibinfo{year}{2017}).

\bibitem{yeh_direct_2016}
\bibinfo{author}{Yeh, P.-C.} \emph{et~al.}
\newblock \bibinfo{journal}{\bibinfo{title}{Direct {Measurement} of the
  {Tunable} {Electronic} {Structure} of {Bilayer} {MoS$_2$} by {Interlayer}
  {Twist}}}.
\newblock {\emph{Nano Letters}} \textbf{\bibinfo{volume}{16}},
  \bibinfo{pages}{953--959}, \doiprefix\url{10.1021/acs.nanolett.5b03883}
  (\bibinfo{year}{2016}).

\bibitem{kumari_engineering_2020}
\bibinfo{author}{Kumari, P.}, \bibinfo{author}{Chatterjee, J.} \&
  \bibinfo{author}{Mahadevan, P.}
\newblock \bibinfo{journal}{\bibinfo{title}{Engineering spin-valley physics in
  bilayers of {MoSe$_2$}}}.
\newblock {\emph{Physical Review B}}
  \textbf{\bibinfo{volume}{101}}, \bibinfo{pages}{045432},
  \doiprefix\url{10.1103/PhysRevB.101.045432} (\bibinfo{year}{2020}).

\bibitem{scuri_electrically_2020}
\bibinfo{author}{Scuri, G.} \emph{et~al.}
\newblock \bibinfo{journal}{\bibinfo{title}{Electrically {Tunable} {Valley}
  {Dynamics} in {Twisted} wse$_2$/wse$_2$ {Bilayers}}}.
\newblock {\emph{Physical Review Letters}}
  \textbf{\bibinfo{volume}{124}}, \bibinfo{pages}{217403},
  \doiprefix\url{10.1103/PhysRevLett.124.217403} (\bibinfo{year}{2020}).

\bibitem{zhang_flat_2020}
\bibinfo{author}{Zhang, Z.} \emph{et~al.}
\newblock \bibinfo{journal}{\bibinfo{title}{Flat bands in twisted bilayer
  transition metal dichalcogenides}}.
\newblock {\emph{Nature Physics}} \bibinfo{pages}{1--4},
  \doiprefix\url{10.1038/s41567-020-0958-x} (\bibinfo{year}{2020}).

\bibitem{novoselov_nobel_2011}
\bibinfo{author}{Novoselov, K.~S.}
\newblock \bibinfo{journal}{\bibinfo{title}{Nobel {Lecture}: {Graphene}:
  {Materials} in the {Flatland}}}.
\newblock {\emph{Reviews of Modern Physics}}
  \textbf{\bibinfo{volume}{83}}, \bibinfo{pages}{837--849},
  \doiprefix\url{10.1103/RevModPhys.83.837} (\bibinfo{year}{2011}).

\bibitem{kang_high-mobility_2015}
\bibinfo{author}{Kang, K.} \emph{et~al.}
\newblock \bibinfo{journal}{\bibinfo{title}{High-mobility three-atom-thick
  semiconducting films with wafer-scale homogeneity}}.
\newblock {\emph{Nature}} \textbf{\bibinfo{volume}{520}},
  \bibinfo{pages}{656--660}, \doiprefix\url{10.1038/nature14417}
  (\bibinfo{year}{2015}).

\bibitem{wang_dual-coupling-guided_2022}
\bibinfo{author}{Wang, J.} \emph{et~al.}
\newblock \bibinfo{journal}{\bibinfo{title}{Dual-coupling-guided epitaxial
  growth of wafer-scale single-crystal {WS$_2$} monolayer on vicinal a-plane
  sapphire}}.
\newblock {\emph{Nature Nanotechnology}}
  \textbf{\bibinfo{volume}{17}}, \bibinfo{pages}{33--38},
  \doiprefix\url{10.1038/s41565-021-01004-0} (\bibinfo{year}{2022}).

\bibitem{dau_beyond_2018}
\bibinfo{author}{Dau, M.~T.} \emph{et~al.}
\newblock \bibinfo{journal}{\bibinfo{title}{Beyond van der {Waals}
  {Interaction}: {The} {Case} of {MoSe$_2$} {Epitaxially} {Grown} on
  {Few}-{Layer} {Graphene}}}.
\newblock {\emph{ACS Nano}} \textbf{\bibinfo{volume}{12}},
  \bibinfo{pages}{2319--2331}, \doiprefix\url{10.1021/acsnano.7b07446}
  (\bibinfo{year}{2018}).

\bibitem{pierucci_evidence_2022}
\bibinfo{author}{Pierucci, D.} \emph{et~al.}
\newblock \bibinfo{journal}{\bibinfo{title}{Evidence for highly p-type doping
  and type {II} band alignment in large scale monolayer
  {WSe$_2$}/{Se}-terminated {GaAs} heterojunction grown by molecular beam
  epitaxy}}.
\newblock {\emph{Nanoscale}} \textbf{\bibinfo{volume}{14}},
  \bibinfo{pages}{5859--5868}, \doiprefix\url{10.1039/D2NR00458E}
  (\bibinfo{year}{2022}).

\bibitem{lipatov_nanodomain_2019}
\bibinfo{author}{Lipatov, A.}, \bibinfo{author}{Li, T.},
  \bibinfo{author}{Vorobeva, N.~S.}, \bibinfo{author}{Sinitskii, A.} \&
  \bibinfo{author}{Gruverman, A.}
\newblock \bibinfo{journal}{\bibinfo{title}{Nanodomain {Engineering} for
  {Programmable} {Ferroelectric} {Devices}}}.
\newblock {\emph{Nano Letters}} \textbf{\bibinfo{volume}{19}},
  \bibinfo{pages}{3194--3198}, \doiprefix\url{10.1021/acs.nanolett.9b00673}
  (\bibinfo{year}{2019}).

\bibitem{li_polarization-mediated_2017}
\bibinfo{author}{Li, T.} \emph{et~al.}
\newblock \bibinfo{journal}{\bibinfo{title}{Polarization-{Mediated}
  {Modulation} of {Electronic} and {Transport} {Properties} of {Hybrid}
  {MoS$_2$}-{BaTiO$_3$}–{SrRuO$_3$} {Tunnel} {Junctions}}}.
\newblock {\emph{Nano Letters}} \textbf{\bibinfo{volume}{17}},
  \bibinfo{pages}{922--927}, \doiprefix\url{10.1021/acs.nanolett.6b04247}
  (\bibinfo{year}{2017}).

\bibitem{li_optical_2018}
\bibinfo{author}{Li, T.} \emph{et~al.}
\newblock \bibinfo{journal}{\bibinfo{title}{Optical control of polarization in
  ferroelectric heterostructures}}.
\newblock {\emph{Nature Communications}}
  \textbf{\bibinfo{volume}{9}}, \bibinfo{pages}{1--8},
  \doiprefix\url{10.1038/s41467-018-05640-4} (\bibinfo{year}{2018}).

\bibitem{silva_resistive_2017}
\bibinfo{author}{Silva, J. P.~B.}, \bibinfo{author}{Marques, C.~A.},
  \bibinfo{author}{Moreira, J.~A.} \& \bibinfo{author}{Conde, O.}
\newblock \bibinfo{journal}{\bibinfo{title}{Resistive switching in
  {MoSe$_2$}/{BaTiO$_3$} hybrid structures}}.
\newblock {\emph{Journal of Materials Chemistry C}}
  \textbf{\bibinfo{volume}{5}}, \bibinfo{pages}{10353--10359},
  \doiprefix\url{10.1039/C7TC03024J} (\bibinfo{year}{2017}).

\bibitem{wen_ferroelectric-driven_2019}
\bibinfo{author}{Wen, B.} \emph{et~al.}
\newblock \bibinfo{journal}{\bibinfo{title}{Ferroelectric-{Driven} {Exciton}
  and {Trion} {Modulation} in {Monolayer} {Molybdenum} and {Tungsten}
  {Diselenides}}}.
\newblock {\emph{ACS Nano}} \textbf{\bibinfo{volume}{13}},
  \bibinfo{pages}{5335--5343}, \doiprefix\url{10.1021/acsnano.8b09800}
  (\bibinfo{year}{2019}).

\bibitem{dau_van_2019}
\bibinfo{author}{Dau, M.~T.} \emph{et~al.}
\newblock \bibinfo{journal}{\bibinfo{title}{van der {Waals} epitaxy of
  {Mn}-doped {MoSe$_2$} on mica}}.
\newblock {\emph{APL Materials}} \textbf{\bibinfo{volume}{7}},
  \bibinfo{pages}{051111}, \doiprefix\url{10.1063/1.5093384}
  (\bibinfo{year}{2019}).

\bibitem{yu_interface_2012}
\bibinfo{author}{Yu, P.} \emph{et~al.}
\newblock \bibinfo{journal}{\bibinfo{title}{Interface control of bulk
  ferroelectric polarization}}.
\newblock {\emph{Proceedings of the National Academy of
  Sciences}} \textbf{\bibinfo{volume}{109}}, \bibinfo{pages}{9710--9715},
  \doiprefix\url{10.1073/pnas.1117990109} (\bibinfo{year}{2012}).

\bibitem{vergnaud_new_2020}
\bibinfo{author}{Vergnaud, C.} \emph{et~al.}
\newblock \bibinfo{journal}{\bibinfo{title}{New approach for the molecular beam
  epitaxy growth of scalable {WSe$_2$} monolayers}}.
\newblock {\emph{Nanotechnology}} \textbf{\bibinfo{volume}{31}},
  \bibinfo{pages}{255602}, \doiprefix\url{10.1088/1361-6528/ab80fe}
  (\bibinfo{year}{2020}).

\bibitem{refXPS}
\bibinfo{author}{Shallenberger, J.~R.}
\newblock \bibinfo{journal}{\bibinfo{title}{2d tungsten diselenide analyzed by
  xps}}.
\newblock {\emph{Surface Science Spectra}}
  \textbf{\bibinfo{volume}{25}}, \bibinfo{pages}{014001},
  \doiprefix\url{10.1116/1.5016189} (\bibinfo{year}{2018}).
\newblock \eprint{https://doi.org/10.1116/1.5016189}.

\bibitem{nguyen_visualizing_2019}
\bibinfo{author}{Nguyen, P.~V.} \emph{et~al.}
\newblock \bibinfo{journal}{\bibinfo{title}{Visualizing electrostatic gating
  effects in two-dimensional heterostructures}}.
\newblock {\emph{Nature}} \textbf{\bibinfo{volume}{572}},
  \bibinfo{pages}{220--223}, \doiprefix\url{10.1038/s41586-019-1402-1}
  (\bibinfo{year}{2019}).

\bibitem{zhang_electronic_2016}
\bibinfo{author}{Zhang, Y.} \emph{et~al.}
\newblock \bibinfo{journal}{\bibinfo{title}{Electronic {Structure}, {Surface}
  {Doping}, and {Optical} {Response} in {Epitaxial} {WSe$_2$} {Thin} {Films}}}.
\newblock {\emph{Nano Letters}} \textbf{\bibinfo{volume}{16}},
  \bibinfo{pages}{2485--2491}, \doiprefix\url{10.1021/acs.nanolett.6b00059}
  (\bibinfo{year}{2016}).

\bibitem{nakamura_spin_2020}
\bibinfo{author}{Nakamura, H.} \emph{et~al.}
\newblock \bibinfo{journal}{\bibinfo{title}{Spin splitting and strain in
  epitaxial monolayer {WSe$_2$} on graphene}}.
\newblock {\emph{Physical Review B}}
  \textbf{\bibinfo{volume}{101}}, \bibinfo{pages}{165103},
  \doiprefix\url{10.1103/PhysRevB.101.165103} (\bibinfo{year}{2020}).

\bibitem{riley_direct_2014}
\bibinfo{author}{Riley, J.~M.} \emph{et~al.}
\newblock \bibinfo{journal}{\bibinfo{title}{Direct observation of
  spin-polarized bulk bands in an inversion-symmetric semiconductor}}.
\newblock {\emph{Nature Physics}} \textbf{\bibinfo{volume}{10}},
  \bibinfo{pages}{835--839}, \doiprefix\url{10.1038/nphys3105}
  (\bibinfo{year}{2014}).

\bibitem{dai_large_2019}
\bibinfo{author}{Dai, J.-Q.}, \bibinfo{author}{Wang, X.-W.} \&
  \bibinfo{author}{Cao, T.-F.}
\newblock \bibinfo{journal}{\bibinfo{title}{Large {Band} {Offset} in
  {Monolayer} {MoS$_2$} on {Oppositely} {Polarized} {BiFeO$_3$}(0001) {Polar}
  {Surfaces}}}.
\newblock {\emph{The Journal of Physical Chemistry C}}
  \textbf{\bibinfo{volume}{123}}, \bibinfo{pages}{3039--3047},
  \doiprefix\url{10.1021/acs.jpcc.8b12207} (\bibinfo{year}{2019}).

\bibitem{nguyen_toward_2015}
\bibinfo{author}{Nguyen, A.} \emph{et~al.}
\newblock \bibinfo{journal}{\bibinfo{title}{Toward {Ferroelectric} {Control} of
  {Monolayer} {MoS2}}}.
\newblock {\emph{Nano Letters}} \textbf{\bibinfo{volume}{15}},
  \bibinfo{pages}{3364--3369}, \doiprefix\url{10.1021/acs.nanolett.5b00687}
  (\bibinfo{year}{2015}).

\bibitem{zhang_precise_2011}
\bibinfo{author}{Zhang, P.} \emph{et~al.}
\newblock \bibinfo{journal}{\bibinfo{title}{A precise method for visualizing
  dispersive features in image plots}}.
\newblock {\emph{Review of Scientific Instruments}}
  \textbf{\bibinfo{volume}{82}}, \bibinfo{pages}{043712},
  \doiprefix\url{10.1063/1.3585113} (\bibinfo{year}{2011}).

\end{thebibliography}
\end{document}